\documentclass[10pt,a4paper,onecolumn]{article}
\usepackage{marginnote}
\usepackage{graphicx}
\usepackage{xcolor}
\usepackage{authblk,etoolbox}
\usepackage{titlesec}
\usepackage{calc}
\usepackage{tikz}
\usepackage{hyperref}
\hypersetup{colorlinks=true,breaklinks=true,
            urlcolor=[rgb]{0.0, 0.5, 1.0},
            linkcolor=[rgb]{0.0, 0.5, 1.0}
            }
\usepackage{caption}
\usepackage{orcidlink}
\usepackage{tcolorbox}
\usepackage{amssymb,amsmath}
\usepackage{ifxetex,ifluatex}
\usepackage{seqsplit}
\usepackage{xstring}

\usepackage{float}
\let\origfigure\figure
\let\endorigfigure\endfigure
\renewenvironment{figure}[1][2] {
    \expandafter\origfigure\expandafter[H]
} {
    \endorigfigure
}

\usepackage{fixltx2e} 
\usepackage[
  backend=biber,
]{biblatex}
\bibliography{paper.bib}

\let\textttOrig=\texttt
\def\texttt#1{\expandafter\textttOrig{\seqsplit{#1}}}
\renewcommand{\seqinsert}{\ifmmode
  \allowbreak
  \else\penalty6000\hspace{0pt plus 0.02em}\fi}

\makeatletter
\let\href@Orig=\href
\def\href@Urllike#1#2{\href@Orig{#1}{\begingroup
    \def\Url@String{#2}\Url@FormatString
    \endgroup}}
\def\href@Notdoi#1#2{\def\tempa{#1}\def\tempb{#2}%
  \ifx\tempa\tempb\relax\href@Urllike{#1}{#2}\else
  \href@Orig{#1}{#2}\fi}
\def\href#1#2{%
  \IfBeginWith{#1}{https://doi.org}%
  {\href@Urllike{#1}{#2}}{\href@Notdoi{#1}{#2}}}
\makeatother

\newlength{\cslhangindent}
\newlength{\csllabelwidth}
\newenvironment{CSLReferences}[3] 
 {
  \setlength{\parindent}{0pt}
  \ifodd #1 \everypar{\setlength{\hangindent}{\cslhangindent}}\ignorespaces\fi
  \ifnum #2 > 0
  \setlength{\parskip}{#2\baselineskip}
  \fi
 }%
 {}
\usepackage{calc}

\usepackage[top=3.5cm, bottom=3cm, right=1.5cm, left=1.0cm,
            headheight=2.2cm, reversemp, includemp, marginparwidth=4.5cm]{geometry}

\titleformat{\section}
  {\normalfont\sffamily\Large\bfseries}
  {}{0pt}{}
\titleformat{\subsection}
  {\normalfont\sffamily\large\bfseries}
  {}{0pt}{}
\titleformat{\subsubsection}
  {\normalfont\sffamily\bfseries}
  {}{0pt}{}
\titleformat*{\paragraph}
  {\sffamily\normalsize}

\usepackage{fancyhdr}
\makeatletter
\let\ps@plain\ps@fancy
\fancyheadoffset[L]{4.5cm}
\fancyfootoffset[L]{4.5cm}

\definecolor{linky}{rgb}{0.0, 0.5, 1.0}

\newtcolorbox{repobox}
   {colback=red, colframe=red!75!black,
     boxrule=0.5pt, arc=2pt, left=6pt, right=6pt, top=3pt, bottom=3pt}

\newcommand{\ExternalLink}{%
   \tikz[x=1.2ex, y=1.2ex, baseline=-0.05ex]{%
       \begin{scope}[x=1ex, y=1ex]
           \clip (-0.1,-0.1)
               --++ (-0, 1.2)
               --++ (0.6, 0)
               --++ (0, -0.6)
               --++ (0.6, 0)
               --++ (0, -1);
           \path[draw,
               line width = 0.5,
               rounded corners=0.5]
               (0,0) rectangle (1,1);
       \end{scope}
       \path[draw, line width = 0.5] (0.5, 0.5)
           -- (1, 1);
       \path[draw, line width = 0.5] (0.6, 1)
           -- (1, 1) -- (1, 0.6);
       }
   }

\patchcmd{\@maketitle}{center}{flushleft}{}{}
\patchcmd{\@maketitle}{center}{flushleft}{}{}
\patchcmd{\@maketitle}{\LARGE}{\LARGE\sffamily}{}{}
\def\maketitle{{%
  
  \AB@maketitle}}
\makeatletter
\renewcommand\AB@affilsepx{ \protect\Affilfont}
\renewcommand\AB@affilnote[1]{{\bfseries #1}\hspace{3pt}}
\renewcommand{\affil}[2][]%
   {\newaffiltrue\let\AB@blk@and\AB@pand
      \if\relax#1\relax\def\AB@note{\AB@thenote}\else\def\AB@note{#1}%
        \setcounter{Maxaffil}{0}\fi
        \begingroup
        \let\href=\href@Orig
        \let\texttt=\textttOrig
        \let\protect\@unexpandable@protect
        \def\thanks{\protect\thanks}\def\footnote{\protect\footnote}%
        \@temptokena=\expandafter{\AB@authors}%
        {\def\\{\protect\\\protect\Affilfont}\xdef\AB@temp{#2}}%
         \xdef\AB@authors{\the\@temptokena\AB@las\AB@au@str
         \protect\\[\affilsep]\protect\Affilfont\AB@temp}%
         \gdef\AB@las{}\gdef\AB@au@str{}%
        {\def\\{, \ignorespaces}\xdef\AB@temp{#2}}%
        \@temptokena=\expandafter{\AB@affillist}%
        \xdef\AB@affillist{\the\@temptokena \AB@affilsep
          \AB@affilnote{\AB@note}\protect\Affilfont\AB@temp}%
      \endgroup
       \let\AB@affilsep\AB@affilsepx
}
\makeatother

\renewcommand\Affilfont{\sffamily\small\mdseries}
\ifnum 0\ifxetex 1\fi\ifluatex 1\fi=0 
  \usepackage[T1]{fontenc}
  \usepackage[utf8]{inputenc}

\else 
  \ifxetex
    \usepackage{mathspec}
    \usepackage{fontspec}

  \else
    \usepackage{fontspec}
  \fi
  \defaultfontfeatures{Ligatures=TeX,Scale=MatchLowercase}

\fi
\IfFileExists{upquote.sty}{\usepackage{upquote}}{}
\IfFileExists{microtype.sty}{%
\usepackage{microtype}
\UseMicrotypeSet[protrusion]{basicmath} 
}{}

\usepackage{hyperref}
\hypersetup{unicode=true,
            pdftitle={PACMAN: A pipeline to reduce and analyze Hubble Wide Field Camera 3 IR Grism data},
            pdfborder={0 0 0},
            breaklinks=true,
            colorlinks=true,
            linkcolor=[rgb]{0.0, 0.5, 1.0},
            citecolor=Blue,
            urlcolor=[rgb]{0.0, 0.5, 1.0},
            citecolor=Blue}
\let\addcontentslineOrig=\addcontentsline
\def\addcontentsline#1#2#3{\bgroup
  \let\texttt=\textttOrig\addcontentslineOrig{#1}{#2}{#3}\egroup}
\let\markbothOrig\markboth
\def\markboth#1#2{\bgroup
  \let\texttt=\textttOrig\markbothOrig{#1}{#2}\egroup}
\let\markrightOrig\markright
\def\markright#1{\bgroup
  \let\texttt=\textttOrig\markrightOrig{#1}\egroup}

\usepackage{graphicx,grffile}
\makeatletter
\def\maxwidth{\ifdim\Gin@nat@width>\linewidth\linewidth\else\Gin@nat@width\fi}
\def\maxheight{\ifdim\Gin@nat@height>\textheight\textheight\else\Gin@nat@height\fi}
\makeatother
\setkeys{Gin}{width=\maxwidth,height=\maxheight,keepaspectratio}
\IfFileExists{parskip.sty}{%
\usepackage{parskip}
}{
\setlength{\parindent}{0pt}
\setlength{\parskip}{6pt plus 2pt minus 1pt}
}
\providecommand{\tightlist}{%
  \setlength{\itemsep}{0pt}\setlength{\parskip}{0pt}}
\ifx\paragraph\undefined\else
\let\oldparagraph\paragraph
\renewcommand{\paragraph}[1]{\oldparagraph{#1}\mbox{}}
\fi
\ifx\subparagraph\undefined\else
\let\oldsubparagraph\subparagraph
\renewcommand{\subparagraph}[1]{\oldsubparagraph{#1}\mbox{}}
\fi

\title{\texttt{PACMAN}: A pipeline to reduce and analyze Hubble Wide
Field Camera 3 IR Grism data}

        \author[1, 2]{Sebastian Zieba\,\orcidlink{0000-0003-0562-6750}\,}
          \author[1]{Laura Kreidberg\,\orcidlink{0000-0003-0514-1147}\,}
    
      \affil[1]{Max-Planck-Institut für Astronomie, Königstuhl 17,
D-69117 Heidelberg, Germany}
      \affil[2]{Leiden Observatory, Leiden University, Niels Bohrweg 2,
2333CA Leiden, The Netherlands}
  \date{\vspace{-5ex}}

\begin{document}
\maketitle

\marginpar{

  \begin{flushleft}
  \sffamily\small

  {\bfseries DOI:} \href{https://doi.org/10.21105/joss.04838}{\color{linky}{10.21105/joss.04838}}

  \vspace{2mm}

  {\bfseries Software}
  \begin{itemize}
    \setlength\itemsep{0em}
    \item \href{https://github.com/openjournals/joss-reviews/issues/4838}{\color{linky}{Review}} \ExternalLink
    \item \href{https://github.com/sebastian-zieba/PACMAN}{\color{linky}{Repository}} \ExternalLink
    \item \href{https://doi.org/10.5281/zenodo.7465579}{\color{linky}{Archive}} \ExternalLink
  \end{itemize}

  \vspace{2mm}

  \par\noindent\hrulefill\par

  \vspace{2mm}

  {\bfseries Editor:} \href{https://juanjobazan.com/}{Juanjo
Bazán} \ExternalLink \,\orcidlink{0000-0001-7699-3983}\,\\
  \vspace{1mm}
    {\bfseries Reviewers:}
  \begin{itemize}
  \setlength\itemsep{0em}
    \item \href{https://github.com/astrobel}{@astrobel}
    \item \href{https://github.com/aureliocarnero}{@aureliocarnero}
    \end{itemize}
    \vspace{2mm}

  {\bfseries Submitted:} 25 July 2022\\
  {\bfseries Published:} 21 December 2022

  \vspace{2mm}
  {\bfseries License}\\
  Authors of papers retain copyright and release the work under a Creative Commons Attribution 4.0 International License (\href{http://creativecommons.org/licenses/by/4.0/}{\color{linky}{CC BY 4.0}}).

  \end{flushleft}
}

\newcommand{\lktwo}{Kreidberg, Bean, Désert, Line, et al., (\hyperlink{ref-Kreidberg2014}{2014})}

\hypertarget{summary}{%
\section{Summary}\label{summary}}

The Hubble Space Telescope (HST) has become the preeminent workhorse
facility for the characterization of extrasolar planets. Launched in
1990 and never designed for the observations of exoplanets, the STIS
spectrograph on HST was used in 2002 to detect the first atmosphere ever
discovered on a planet outside of our solar system (\hyperlink{ref-Charbonneau2002}{Charbonneau et al.,
2002}).

HST currently has two of the most powerful space-based tools for
characterizing exoplanets over a broad spectral range: The Space
Telescope Imaging Spectrograph (STIS; installed in 1997) in the UV and
the Wide Field Camera 3 (WFC3; installed in 2009) in the Near Infrared
(NIR). With the introduction of a spatial scan mode on WFC3 (\hyperlink{ref-Deming2012}{Deming et al., 2012}; \hyperlink{ref-McCullough2012}{McCullough \& MacKenty, 2012}) where the star moves
perpendicular to the dispersion direction during an exposure, WFC3
observations have become very efficient due to the reduction of overhead
time and the possibility of longer exposures without saturation.

For exoplanet characterization, WFC3 is used for transit and secondary
eclipse spectroscopy, and phase curve observations. The instrument has
two different grisms: G102 with a spectral range from 800 nm to up to
1150 nm and G141 encompassing 1075 nm to about 1700 nm. The spectral
range of WFC3/G141 is primarily sensitive to molecular absorption from
water at approximately 1.4 microns. This led to the successful detection
of water in the atmosphere of over a dozen of exoplanets (e.g., \hyperlink{ref-Deming2013}{Deming
et al., 2013}; \hyperlink{ref-Evans2016}{Evans et al., 2016}; \hyperlink{ref-Fraine2014}{Fraine et al., 2014}; \hyperlink{ref-Huitson2013}{Huitson et al.,
2013}; \hyperlink{ref-Kreidberg2014}{Kreidberg, Bean, Désert, Line, et al., 2014}). The bluer part of
WFC3, the G102 grism, is also sensitive to water and most notably led to
the first detection of a helium exosphere (\hyperlink{ref-Spake2018}{Spake et al., 2018}).

Here we present \texttt{PACMAN}, an end-to-end pipeline developed to
reduce and analyze HST/WFC3 data. The pipeline includes both spectral
extraction and light curve fitting. The foundation of \texttt{PACMAN}
has been already used in numerous publications (e.g., \hyperlink{ref-Kreidberg2014}{Kreidberg, Bean, Désert, Line, et al., 2014}; \hyperlink{Kreidberg2018}{Kreidberg et al., 2018}) and these papers
have already accumulated hundreds of citations.

\hypertarget{statement-of-need}{%
\section{Statement of need}\label{statement-of-need}}

Exoplanet spectroscopy with Hubble requires very precise measurements
that are beyond the scope of standard analysis tools provided by the
Space Telescope Science Institute. The data analysis is challenging, and
different pipelines have produced discrepant results in the literature
(e.g., \hyperlink{ref-Kreidberg2019}{Kreidberg et al., 2019}; \hyperlink{Teachey2018}{Teachey \& Kipping, 2018}). To facilitate
reproducibility and transparency, the data reduction and analysis
software should be open-source. This will enable easy comparison between
different pipelines, and also lower the barrier to entry for newcomers
in the exoplanet atmosphere field.

What sets \texttt{PACMAN} apart from other tools provided by the
community, is that it was specifically designed to reduce and fit HST
data. There are several open-source tools that can fit time series
observations of stars to model events like transiting exoplanets, such
as \texttt{EXOFASTv2} (\hyperlink{ref-Eastman2019}{Eastman et al., 2019}), \texttt{juliet} (\hyperlink{ref-Espinoza2019}{Espinoza
et al., 2019}), \texttt{allesfitter} (\hyperlink{ref-Gunther2019}{Günther \& Daylan, 2019} \hyperlink{ref-Gunther2021}{, 2021}),
\texttt{exoplanet} (\hyperlink{ref-Foreman-Mackey2021a}{Foreman-Mackey et al., 2021a,} \hyperlink{ref-Foreman-Mackey2021b}{2021b}), and
\texttt{starry} (\hyperlink{Luger2019}{Luger et al., 2019}). \texttt{PACMAN}'s source code,
however, includes fitting models that can model systematics which are
characteristic to HST data, such as the orbit-long exponential ramps due
to charge trapping or the upstream-downstream effect. This removes the
need for the user to write these functions themselves. \texttt{PACMAN}
will also retrieve information from the header of the FITS files,
automatically detect HST orbits and visits and use this information in
the fitting models.

The only other end-to-end open source pipeline specifically developed
for the reduction and analysis of HST/WFC3 data is
\href{https://github.com/ucl-exoplanets/Iraclis}{\texttt{Iraclis}}
(\hyperlink{ref-Tsiaras2016}{Tsiaras et al., 2016}). Another open-source pipeline that has been for
example used as an independent check of recent results presented in Mugnai et al. (\hyperlink{Mugnai2021}{2021}) and Carone et al. (\hyperlink{ref-Carone2021}{2021}) is
\href{https://jbouwman.gitlab.io/CASCADe/}{\texttt{CASCADe}}
(Calibration of trAnsit Spectroscopy using CAusal Data). For a more
detailed discussion of \texttt{CASCADe} see Appendix 1 in Carone et al. (\hyperlink{ref-Carone2021}{2021}).

\hypertarget{outline-of-the-pipeline-steps}{%
\section{Outline of the pipeline
steps}\label{outline-of-the-pipeline-steps}}

The pipeline starts with the \emph{ima} data products provided by the
Space Telescope Science Institute that can be easily accessed from
\href{https://mast.stsci.edu/search/hst}{MAST}. These files created by
the WFC3 calibration pipeline, \texttt{calwf3}, have already several
calibrations applied (dark subtraction, linearity correction,
flat-fielding) to each readout of the IR exposure.

\begin{figure}
\centering
\includegraphics[width=0.999\textwidth,height=\textheight]{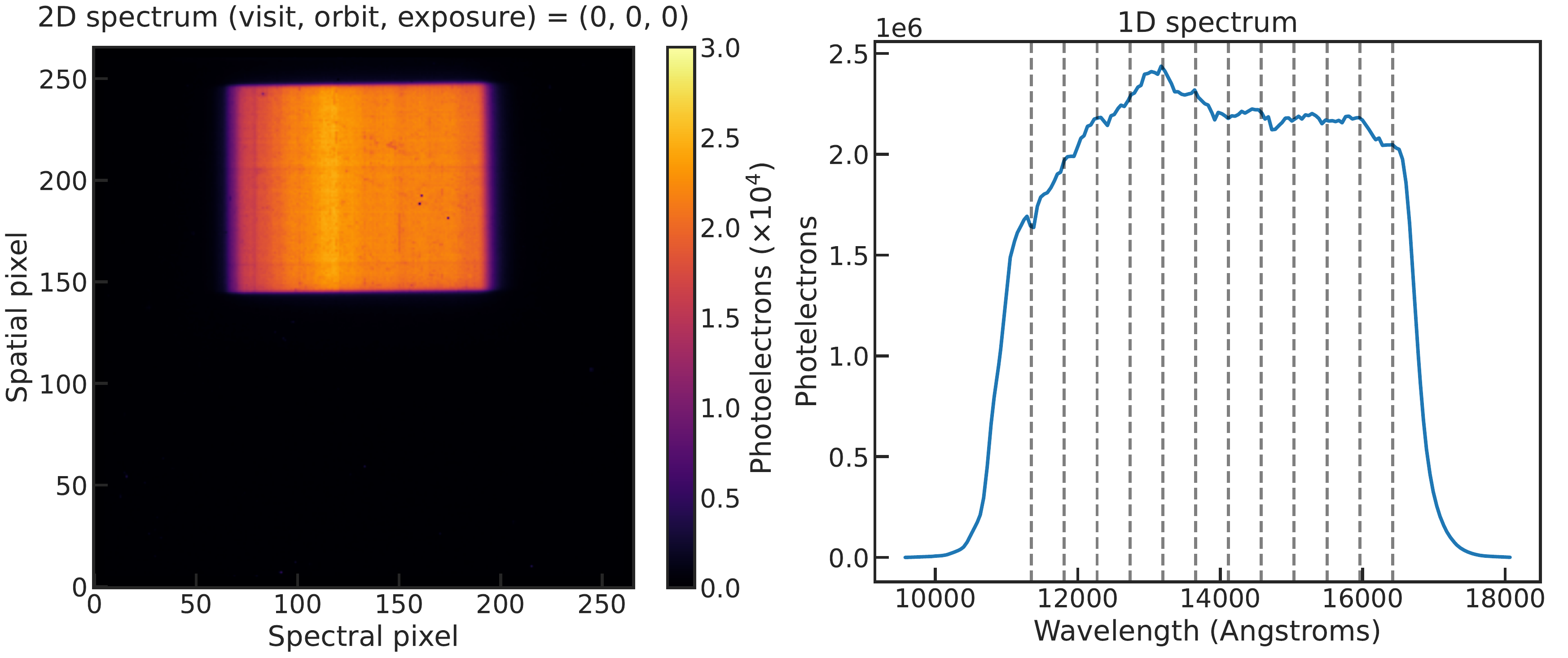}
\caption{\emph{Left panel}: a typical single exposure showing the raw 2D
spectrum. \emph{Right panel}: 1D spectrum after the use of optimal
extraction including vertical dashed lines showing the user-set binning
to generate spectroscopic light curves.\label{fig:figure1}}
\end{figure}

\begin{figure}
\centering
\includegraphics[width=0.999\textwidth,height=\textheight]{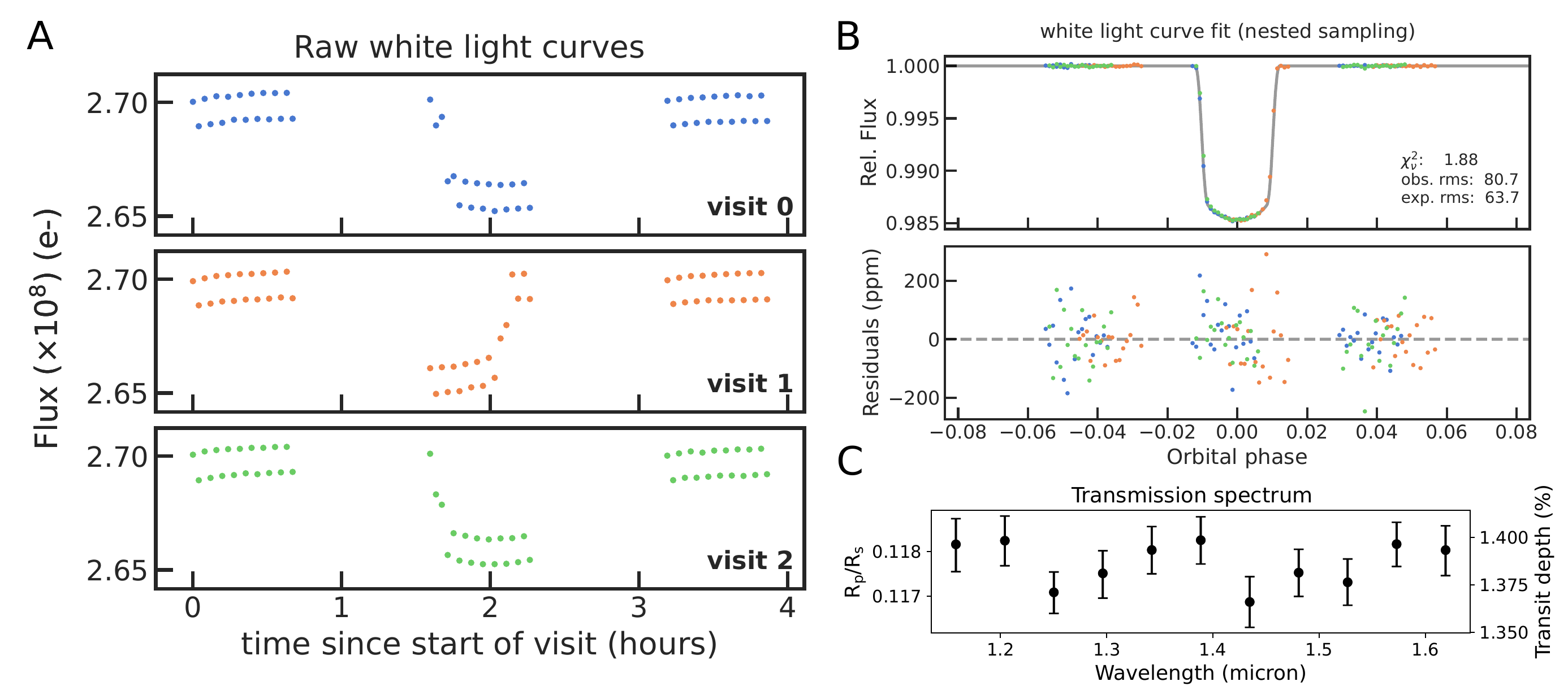}
\caption{\emph{panel A}: raw white light curves for each of the three
visits. One can clearly see the constant offset between two adjacent
exposures due to the spatial scanning mode. \emph{panel B}: white light
curve with the best astrophysical model fit using
\autoref{eq:equation1}. \emph{panel C}: the transmission spectrum after
fitting 11 spectroscopic light curves revealing the flat spectrum of GJ 1214 b as published in \protect\lktwo. \label{fig:figure2}}
\end{figure}

In the following we highlight several steps in the reduction and fitting
stages of the code which are typical for HST/WFC3 observations:

\begin{itemize}
\item
  \textbf{Wavelength calibration}: We create a reference spectrum based
  on the throughput of the respective grism (G102 or G141) and a stellar
  model. The user can decide if he or she wants to download a stellar
  spectrum from MAST or use a black body spectrum. This template is used
  for the wavelength calibration of the WFC3 spectra. We also determine
  the position of the star in the direct images which are commonly taken
  at the start of HST orbits to create an initial guess for the
  wavelength solution using the known dispersion of the grism. Using the
  reference spectrum as a template, we determine a shift and scaling in
  wavelength-space that minimizes the difference between the template
  and the first spectrum in the visit. This first exposure in the visit
  is then used as the template for the following exposures in the visit.
\item
  \textbf{Optimal extraction and outlier removal}: \texttt{PACMAN} uses
  an optimal extraction algorithm as presented in Horne (\hyperlink{ref-Horne1986}{1986}) which
  iteratively masks bad pixels in the image. We also mask bad pixels
  that have been flagged by \texttt{calwf3} with data quality DQ = 4 or
  512\footnote{for a list of DQ flags see \url{https://wfc3tools.readthedocs.io/en/latest/wfc3tools/calwf3.html}}.
\item
  \textbf{Scanning of the detector}: The majority of exoplanetary
  HST/WFC3 observations use the spatial scanning technique (\hyperlink{ref-McCullough2012}{McCullough \& MacKenty, 2012}) which spreads the light perpendicular to the
  dispersion direction during the exposure enabling longer integration
  times before saturation. The \emph{ima} files taken in this
  observation mode consist of a number of nondestructive reads, also
  known as up-the-ramp samples, each of which we treat as an independent
  subexposure. \autoref{fig:figure1} (left panel) shows an example of
  the last subexposure when using spatial scanning together with the
  expected position of the trace based on the direct image.
\item
  \textbf{Fitting models}: \texttt{PACMAN} contains several functions to
  fit models which are commonly used with HST data. The user can fit
  models like in \autoref{eq:equation1} to the white light curve or to
  spectroscopic light curves. An example of a raw spectroscopic light
  curve and fitting \autoref{eq:equation1} to it, can be found in
  \autoref{fig:figure2}. Here are some examples of the currently
  implemented models for the instrument systematics and the
  astrophysical signal:

  \begin{itemize}
  \tightlist
  \item
    systematics models:

    \begin{itemize}
    \tightlist
    \item
      visit-long polynomials
    \item
      orbit-long exponential ramps due to charge trapping: NIR detectors
      like HST/WFC3 can trap photoelectrons (\hyperlink{ref-Smith2008}{Smith et al., 2008}), which
      will cause the number of recorded photoelectrons to increase
      exponentially, creating typical hook-like features in each orbit
    \end{itemize}
  \item
    astrophysical models:

    \begin{itemize}
    \tightlist
    \item
      transit and secondary eclipse curves as implemented in
      \texttt{batman}
    \item
      sinusoids for phase curve fits
    \item
      a constant offset that accounts for the upstream-downstream effect
      (\hyperlink{ref-McCullough2012}{McCullough \& MacKenty, 2012}) caused by forward and reverse
      scanning
    \end{itemize}
  \end{itemize}

  A typical model to fit an exoplanet transit in HST data is the
  following (used, for example, by \hyperlink{ref-Kreidberg2014}{Kreidberg, Bean, Désert, Line, et al., 2014}):

  \begin{equation}
    \label{eq:equation1}
    F(t) = T(t) \, (c\,S(t) + k\,t_{\rm{v}}) \, (1 - \exp(-r_1\,t_{\rm{orb}} - r_2 )),
    \end{equation}

  with \emph{T(t)} being the transit model, \emph{c} (\emph{k}) a
  constant (slope), \emph{S(t)} a scale factor equal to 1 for exposures
  with spatial scanning in the forward direction, and \emph{s} for
  reverse scans, \(r_{\rm{1}}\) and \(r_{\rm{2}}\) are parameters to
  account for the exponential ramps. \(t_{\rm{v}}\) and \(t_{\rm{orb}}\)
  are the times from the first exposure in the visit and in the orbit,
  respectively.
\item
  \textbf{Parameter estimation}: The user has different options to
  estimate best fitting parameters and their uncertainties:

  \begin{itemize}
  \tightlist
  \item
    least squared: \texttt{scipy.optimize}
  \item
    MCMC: \texttt{emcee} (\hyperlink{ref-Foreman-Mackey2013}{Foreman-Mackey et al., 2013})
  \item
    nested sampling: \texttt{dynesty} (\hyperlink{ref-Speagle2020}{Speagle, 2020})
  \end{itemize}
\item
  \textbf{Multi-visit observations}

  \begin{itemize}
  \tightlist
  \item
    \texttt{PACMAN} has also an option to share parameters across
    visits.
  \end{itemize}
\item
  \textbf{Binning of the light spectrum}: The user can freely specify
  the bin numbers or locations. \autoref{fig:figure1} (right panel)
  shows the resulting 1D spectrum and a user-defined binning.
\end{itemize}

\autoref{fig:figure1} and \autoref{fig:figure2} show some figures
created by \texttt{PACMAN} during a run using three HST visits of GJ
1214 b collected in
\href{https://archive.stsci.edu/proposal_search.php?id=13021\&mission=hst}{GO
13201} (\hyperlink{ref-Bean2012}{Bean, 2012}). An analysis of all 15 visits was published in
\lktwo. The analysis of three
visits here using \texttt{PACMAN}, is consistent with the published
results.

\hypertarget{dependencies}{%
\section{Dependencies}\label{dependencies}}

\texttt{PACMAN} uses typical dependencies of astrophysical Python codes:
\texttt{numpy} (\hyperlink{ref-numpy2020}{Harris et al., 2020}), ~\, \,
\texttt{matplotlib} (\hyperlink{ref-matplotlib2007}{Hunter,
2007}), \texttt{scipy} (\hyperlink{ref-scipy2020}{Virtanen et al., 2020}) and \texttt{astropy}
(\hyperlink{ref-astropy2022}{Astropy Collaboration et al., 2022}, \hyperlink{ref-astropy2018}{2018}, \hyperlink{ref-astropy2013}{2013}).

Other dependencies required for the fitting stage depending on the model
and sampler being run are: \texttt{batman} (\hyperlink{ref-Kreidberg2015}{Kreidberg, 2015}),
\texttt{emcee} (\hyperlink{ref-Foreman-Mackey2013}{Foreman-Mackey et al., 2013}), \texttt{dynesty} (\hyperlink{ref-Speagle2020}{Speagle, 2020}), and \texttt{corner} (\hyperlink{ref-Foreman-Mackey2016}{Foreman-Mackey, 2016}).

For the barycentric correction, \texttt{PACMAN} accesses the
\href{https://ssd-api.jpl.nasa.gov/obsolete/horizons_batch_cgi.html}{API
to JPL's Horizons system}.

If the user decides to use a stellar spectrum for the wavelength
calibration, \texttt{PACMAN} will download the needed fits file from the
\href{https://archive.stsci.edu/hlsps/reference-atlases/cdbs/grid/}{``REFERENCE-ATLASES'\,'
high level science product} hosted on the MAST archive (\hyperlink{ref-STScI2013}{STScI
Development Team, 2013}).

\hypertarget{documentation}{%
\section{Documentation}\label{documentation}}

The documentation for \texttt{PACMAN} can be found at
\href{https://pacmandocs.readthedocs.io/en/latest/}{pacmandocs.readthedocs.io}
hosted on \href{https://readthedocs.org/}{ReadTheDocs}. It includes most
notably, a full explanation of every parameter in the \emph{pacman
control file} (pcf), the API, and an example of how to download, reduce
and analyze observations of GJ 1214 b taken with HST/WFC3/G141.

\hypertarget{future-work}{%
\section{Future work}\label{future-work}}

The following features are planned for future development:

\begin{itemize}
\tightlist
\item
  The addition of fitting models like phase curves using the open-source
  Python package \texttt{SPIDERMAN} (\hyperlink{ref-Louden2018}{Louden \& Kreidberg, 2018}).
\item
  Orbit-long ramp fitting using the
  \href{https://recte.readthedocs.io/en/latest/}{RECTE systematic
  model}.
\item
  Limb darkening calculations for users wanting to fix limb darkening
  parameters to theoretical models in the fitting stage.
\item
  Extension to WFC3/UVIS data reduction.
\end{itemize}

\hypertarget{acknowledgements}{%
\section{Acknowledgements}\label{acknowledgements}}

We acknowledge B. Zawadzki for the creation of the \texttt{PACMAN} logo.
We also acknowledge the comments and contributions by I. Momcheva to
\texttt{PACMAN}.

\hypertarget{references}{%
\section*{References}\label{references}}
\addcontentsline{toc}{section}{References}

\hypertarget{refs}{}
\begin{CSLReferences}{1}{0}
\leavevmode\hypertarget{ref-astropy2022}{}%
Astropy Collaboration, Price-Whelan, A. M., Lim, P. L., Earl, N.,
Starkman, N., Bradley, L., Shupe, D. L., Patil, A. A., Corrales, L.,
Brasseur, C. E., Nöthe, M., Donath, A., Tollerud, E., Morris, B. M.,
Ginsburg, A., Vaher, E., Weaver, B. A., Tocknell, J., Jamieson, W.,
\ldots{} Astropy Project Contributors. (2022). {The Astropy Project:
Sustaining and Growing a Community-oriented Open-source Project and the
Latest Major Release (v5.0) of the Core Package}. \emph{Astrophysical
Journal}, \emph{935}(2), 167.
\url{https://doi.org/10.3847/1538-4357/ac7c74}

\leavevmode\hypertarget{ref-astropy2018}{}%
Astropy Collaboration, Price-Whelan, A. M., Sipőcz, B. M., Günther, H.
M., Lim, P. L., Crawford, S. M., Conseil, S., Shupe, D. L., Craig, M.
W., Dencheva, N., Ginsburg, A., VanderPlas, J. T., Bradley, L. D.,
Pérez-Suárez, D., de Val-Borro, M., Aldcroft, T. L., Cruz, K. L.,
Robitaille, T. P., Tollerud, E. J., \ldots{} Astropy Contributors.
(2018). {The Astropy Project: Building an Open-science Project and
Status of the v2.0 Core Package}. \emph{Astronomical Journal},
\emph{156}(3), 123. \url{https://doi.org/10.3847/1538-3881/aabc4f}

\leavevmode\hypertarget{ref-astropy2013}{}%
Astropy Collaboration, Robitaille, T. P., Tollerud, E. J., Greenfield,
P., Droettboom, M., Bray, E., Aldcroft, T., Davis, M., Ginsburg, A.,
Price-Whelan, A. M., Kerzendorf, W. E., Conley, A., Crighton, N.,
Barbary, K., Muna, D., Ferguson, H., Grollier, F., Parikh, M. M., Nair,
P. H., \ldots{} Streicher, O. (2013). {Astropy: A community Python
package for astronomy}. \emph{Astronomy and Astrophysics}, \emph{558},
A33. \url{https://doi.org/10.1051/0004-6361/201322068}

\leavevmode\hypertarget{ref-Bean2012}{}%
Bean, J. (2012). \emph{{Revealing the Diversity of Super-Earth
Atmospheres}} (p. 13021). HST Proposal ID 13021. Cycle 20.

\leavevmode\hypertarget{ref-Carone2021}{}%
Carone, L., Mollière, P., Zhou, Y., Bouwman, J., Yan, F., Baeyens, R.,
Apai, D., Espinoza, N., Rackham, B. V., Jordán, A., Angerhausen, D.,
Decin, L., Lendl, M., Venot, O., \& Henning, T. (2021). {Indications for
very high metallicity and absence of methane in the eccentric exo-Saturn
WASP-117b}. \emph{Astronomy and Astrophysics}, \emph{646}, A168.
\url{https://doi.org/10.1051/0004-6361/202038620}

\leavevmode\hypertarget{ref-Charbonneau2002}{}%
Charbonneau, D., Brown, T. M., Noyes, R. W., \& Gilliland, R. L. (2002).
{Detection of an Extrasolar Planet Atmosphere}. \emph{Astrophysical
Journal}, \emph{568}(1), 377--384. \url{https://doi.org/10.1086/338770}

\leavevmode\hypertarget{ref-Deming2013}{}%
Deming, D., Wilkins, A., McCullough, P., Burrows, A., Fortney, J. J.,
Agol, E., Dobbs-Dixon, I., Madhusudhan, N., Crouzet, N., Desert, J.-M.,
Gilliland, R. L., Haynes, K., Knutson, H. A., Line, M., Magic, Z.,
Mandell, A. M., Ranjan, S., Charbonneau, D., Clampin, M., \ldots{}
Showman, A. P. (2013). {Infrared Transmission Spectroscopy of the
Exoplanets HD 209458b and XO-1b Using the Wide Field Camera-3 on the
Hubble Space Telescope}. \emph{Astrophysical Journal}, \emph{774}(2),
95. \url{https://doi.org/10.1088/0004-637X/774/2/95}

\leavevmode\hypertarget{ref-Deming2012}{}%
Deming, D., Wilkins, A., McCullough, P., Madhusudhan, N., Agol, E.,
Burrows, A., Charbonneau, D., Clampin, M., Desert, J., Gilliland, R.,
Knutson, H., Mandell, A., Ranjan, S., Seager, S., \& Showman, A. (2012).
{Infrared Spectroscopy of the Transiting Exoplanets HD189733b and XO-1
Using Hubble WFC3 in Spatial Scan Mode}. \emph{American Astronomical
Society Meeting Abstracts \#219}, \emph{219}, 405.05.

\leavevmode\hypertarget{ref-Eastman2019}{}%
Eastman, J. D., Rodriguez, J. E., Agol, E., Stassun, K. G., Beatty, T.
G., Vanderburg, A., Gaudi, B. S., Collins, K. A., \& Luger, R. (2019).
{EXOFASTv2: A public, generalized, publication-quality exoplanet
modeling code}. \emph{arXiv e-Prints}, arXiv:1907.09480.
\url{http://arxiv.org/abs/1907.09480}

\leavevmode\hypertarget{ref-Espinoza2019}{}%
Espinoza, N., Kossakowski, D., \& Brahm, R. (2019). {juliet: a versatile
modelling tool for transiting and non-transiting exoplanetary systems}.
\emph{Monthly Notices of the RAS}, \emph{490}(2), 2262--2283.
\url{https://doi.org/10.1093/mnras/stz2688}

\leavevmode\hypertarget{ref-Evans2016}{}%
Evans, T. M., Sing, D. K., Wakeford, H. R., Nikolov, N., Ballester, G.
E., Drummond, B., Kataria, T., Gibson, N. P., Amundsen, D. S., \& Spake,
J. (2016). {Detection of H\(_{2}\)O and Evidence for TiO/VO in an
Ultra-hot Exoplanet Atmosphere}. \emph{Astrophysical Journal, Letters},
\emph{822}(1), L4. \url{https://doi.org/10.3847/2041-8205/822/1/L4}

\leavevmode\hypertarget{ref-corner2016}{}%
Foreman-Mackey, D. (2016). Corner.py: Scatterplot matrices in python.
\emph{The Journal of Open Source Software}, \emph{1}(2), 24.
\url{https://doi.org/10.21105/joss.00024}

\leavevmode\hypertarget{ref-Foreman-Mackey2013}{}%
Foreman-Mackey, D., Hogg, D. W., Lang, D., \& Goodman, J. (2013).
{emcee: The MCMC Hammer}. \emph{Publications of the ASP},
\emph{125}(925), 306. \url{https://doi.org/10.1086/670067}

\leavevmode\hypertarget{ref-Foreman-Mackey2021a}{}%
Foreman-Mackey, D., Luger, R., Agol, E., Barclay, T., Bouma, L. G.,
Brandt, T. D., Czekala, I., David, T. J., Dong, J., Gilbert, E. A.,
Gordon, T. A., Hedges, C., Hey, D. R., Morris, B. M., Price-Whelan, A.
M., \& Savel, A. B. (2021a). \emph{{exoplanet: Gradient-based
probabilistic inference for exoplanet data \& other astronomical time
series}} (Version 0.5.1). Zenodo; Zenodo.
\url{https://doi.org/10.5281/zenodo.1998447}

\leavevmode\hypertarget{ref-Foreman-Mackey2021b}{}%
Foreman-Mackey, D., Luger, R., Agol, E., Barclay, T., Bouma, L., Brandt,
T., Czekala, I., David, T., Dong, J., Gilbert, E., Gordon, T., Hedges,
C., Hey, D., Morris, B., Price-Whelan, A., \& Savel, A. (2021b).
{exoplanet: Gradient-based probabilistic inference for exoplanet data \&
other astronomical time series}. \emph{The Journal of Open Source
Software}, \emph{6}(62), 3285. \url{https://doi.org/10.21105/joss.03285}

\leavevmode\hypertarget{ref-Fraine2014}{}%
Fraine, J., Deming, D., Benneke, B., Knutson, H., Jordán, A., Espinoza,
N., Madhusudhan, N., Wilkins, A., \& Todorov, K. (2014). {Water vapour
absorption in the clear atmosphere of a Neptune-sized exoplanet}.
\emph{Nature}, \emph{513}(7519), 526--529.
\url{https://doi.org/10.1038/nature13785}

\leavevmode\hypertarget{ref-Gunther2019}{}%
Günther, M. N., \& Daylan, T. (2019). \emph{{allesfitter: Flexible star
and exoplanet inference from photometry and radial velocity}} (p.
ascl:1903.003). Astrophysics Source Code Library, record ascl:1903.003.

\leavevmode\hypertarget{ref-Gunther2021}{}%
Günther, M. N., \& Daylan, T. (2021). {Allesfitter: Flexible Star and
Exoplanet Inference from Photometry and Radial Velocity}.
\emph{Astrophysical Journal, Supplement}, \emph{254}(1), 13.
\url{https://doi.org/10.3847/1538-4365/abe70e}

\leavevmode\hypertarget{ref-numpy2020}{}%
Harris, C. R., Millman, K. J., van der Walt, S. J., Gommers, R.,
Virtanen, P., Cournapeau, D., Wieser, E., Taylor, J., Berg, S., Smith,
N. J., Kern, R., Picus, M., Hoyer, S., van Kerkwijk, M. H., Brett, M.,
Haldane, A., del R{\'{i}}o, J. F., Wiebe, M., Peterson, P., \ldots{} Oliphant,
T. E. (2020). {Array programming with NumPy}. \emph{Nature},
\emph{585}(7825), 357--362.
\url{https://doi.org/10.1038/s41586-020-2649-2}

\leavevmode\hypertarget{ref-Horne1986}{}%
Horne, K. (1986). {An optimal extraction algorithm for CCD
spectroscopy.} \emph{Publications of the ASP}, \emph{98}, 609--617.
\url{https://doi.org/10.1086/131801}

\leavevmode\hypertarget{ref-Huitson2013}{}%
Huitson, C. M., Sing, D. K., Pont, F., Fortney, J. J., Burrows, A. S.,
Wilson, P. A., Ballester, G. E., Nikolov, N., Gibson, N. P., Deming, D.,
Aigrain, S., Evans, T. M., Henry, G. W., Lecavelier des Etangs, A.,
Showman, A. P., Vidal-Madjar, A., \& Zahnle, K. (2013). {An HST
optical-to-near-IR transmission spectrum of the hot Jupiter WASP-19b:
detection of atmospheric water and likely absence of TiO}. \emph{Monthly
Notices of the RAS}, \emph{434}(4), 3252--3274.
\url{https://doi.org/10.1093/mnras/stt1243}

\leavevmode\hypertarget{ref-matplotlib2007}{}%
Hunter, J. D. (2007). {Matplotlib: A 2D Graphics Environment}.
\emph{Computing in Science and Engineering}, \emph{9}(3), 90--95.
\url{https://doi.org/10.1109/MCSE.2007.55}

\leavevmode\hypertarget{ref-Kreidberg2015}{}%
Kreidberg, L. (2015). {batman: BAsic Transit Model cAlculatioN in
Python}. \emph{Publications of the ASP}, \emph{127}(957), 1161.
\url{https://doi.org/10.1086/683602}

\leavevmode\hypertarget{ref-Kreidberg2014a}{}%
Kreidberg, L., Bean, J. L., Désert, J.-M., Benneke, B., Deming, D.,
Stevenson, K. B., Seager, S., Berta-Thompson, Z., Seifahrt, A., \&
Homeier, D. (2014). {Clouds in the atmosphere of the super-Earth
exoplanet GJ1214b}. \emph{Nature}, \emph{505}(7481), 69--72.
\url{https://doi.org/10.1038/nature12888}

\leavevmode\hypertarget{ref-Kreidberg2014b}{}%
Kreidberg, L., Bean, J. L., Désert, J.-M., Line, M. R., Fortney, J. J.,
Madhusudhan, N., Stevenson, K. B., Showman, A. P., Charbonneau, D.,
McCullough, P. R., Seager, S., Burrows, A., Henry, G. W., Williamson,
M., Kataria, T., \& Homeier, D. (2014). {A Precise Water Abundance
Measurement for the Hot Jupiter WASP-43b}. \emph{Astrophysical Journal,
Letters}, \emph{793}(2), L27.
\url{https://doi.org/10.1088/2041-8205/793/2/L27}

\leavevmode\hypertarget{ref-Kreidberg2018}{}%
Kreidberg, L., Line, M. R., Parmentier, V., Stevenson, K. B., Louden,
T., Bonnefoy, M., Faherty, J. K., Henry, G. W., Williamson, M. H.,
Stassun, K., Beatty, T. G., Bean, J. L., Fortney, J. J., Showman, A. P.,
Désert, J.-M., \& Arcangeli, J. (2018). {Global Climate and Atmospheric
Composition of the Ultra-hot Jupiter WASP-103b from HST and Spitzer
Phase Curve Observations}. \emph{Astronomical Journal}, \emph{156}(1),
17. \url{https://doi.org/10.3847/1538-3881/aac3df}

\leavevmode\hypertarget{ref-Kreidberg2019}{}%
Kreidberg, L., Luger, R., \& Bedell, M. (2019). {No Evidence for Lunar
Transit in New Analysis of Hubble Space Telescope Observations of the
Kepler-1625 System}. \emph{Astrophysical Journal, Letters},
\emph{877}(2), L15. \url{https://doi.org/10.3847/2041-8213/ab20c8}

\leavevmode\hypertarget{ref-Louden2018}{}%
Louden, T., \& Kreidberg, L. (2018). {SPIDERMAN: an open-source code to
model phase curves and secondary eclipses}. \emph{Monthly Notices of the
RAS}, \emph{477}(2), 2613--2627.
\url{https://doi.org/10.1093/mnras/sty558}

\leavevmode\hypertarget{ref-Luger2019}{}%
Luger, R., Agol, E., Foreman-Mackey, D., Fleming, D. P., Lustig-Yaeger,
J., \& Deitrick, R. (2019). {starry: Analytic Occultation Light Curves}.
\emph{Astronomical Journal}, \emph{157}(2), 64.
\url{https://doi.org/10.3847/1538-3881/aae8e5}

\leavevmode\hypertarget{ref-McCullough2012}{}%
McCullough, P., \& MacKenty, J. (2012). \emph{{Considerations for using
Spatial Scans with WFC3}} (p. 8). Space Telescope WFC Instrument Science
Report.

\leavevmode\hypertarget{ref-Mugnai2021}{}%
Mugnai, L. V., Modirrousta-Galian, D., Edwards, B., Changeat, Q.,
Bouwman, J., Morello, G., Al-Refaie, A., Baeyens, R., Bieger, M. F.,
Blain, D., Gressier, A., Guilluy, G., Jaziri, Y., Kiefer, F., Morvan,
M., Pluriel, W., Poveda, M., Skaf, N., Whiteford, N., \ldots{} Beaulieu,
J.-P. (2021). {ARES. V. No Evidence For Molecular Absorption in the HST
WFC3 Spectrum of GJ 1132 b}. \emph{Astronomical Journal}, \emph{161}(6),
284. \url{https://doi.org/10.3847/1538-3881/abf3c3}

\leavevmode\hypertarget{ref-Smith2008}{}%
Smith, R. M., Zavodny, M., Rahmer, G., \& Bonati, M. (2008). {A theory
for image persistence in HgCdTe photodiodes}. In D. A. Dorn \& A. D.
Holland (Eds.), \emph{High energy, optical, and infrared detectors for
astronomy III} (Vol. 7021, p. 70210J).
\url{https://doi.org/10.1117/12.789372}

\leavevmode\hypertarget{ref-Spake2018}{}%
Spake, J. J., Sing, D. K., Evans, T. M., Oklopčić, A., Bourrier, V.,
Kreidberg, L., Rackham, B. V., Irwin, J., Ehrenreich, D., Wyttenbach,
A., Wakeford, H. R., Zhou, Y., Chubb, K. L., Nikolov, N., Goyal, J. M.,
Henry, G. W., Williamson, M. H., Blumenthal, S., \ldots{} Madhusudhan,
N. (2018). {Helium in the eroding atmosphere of an exoplanet}.
\emph{Nature}, \emph{557}(7703), 68--70.
\url{https://doi.org/10.1038/s41586-018-0067-5}

\leavevmode\hypertarget{ref-Speagle2020}{}%
Speagle, J. S. (2020). {DYNESTY: a dynamic nested sampling package for
estimating Bayesian posteriors and evidences}. \emph{Monthly Notices of
the RAS}, \emph{493}(3), 3132--3158.
\url{https://doi.org/10.1093/mnras/staa278}

\leavevmode\hypertarget{ref-STScI2013}{}%
STScI Development Team. (2013). \emph{{pysynphot: Synthetic photometry
software package}} (p. ascl:1303.023).

\leavevmode\hypertarget{ref-Teachey2018}{}%
Teachey, A., \& Kipping, D. M. (2018). {Evidence for a large exomoon
orbiting Kepler-1625b}. \emph{Science Advances}, \emph{4}(10), eaav1784.
\url{https://doi.org/10.1126/sciadv.aav1784}

\leavevmode\hypertarget{ref-Tsiaras2016}{}%
Tsiaras, A., Waldmann, I. P., Rocchetto, M., Varley, R., Morello, G.,
Damiano, M., \& Tinetti, G. (2016). {A New Approach to Analyzing HST
Spatial Scans: The Transmission Spectrum of HD 209458 b}.
\emph{Astrophysical Journal}, \emph{832}(2), 202.
\url{https://doi.org/10.3847/0004-637X/832/2/202}

\leavevmode\hypertarget{ref-scipy2020}{}%
Virtanen, P., Gommers, R., Oliphant, T. E., Haberland, M., Reddy, T.,
Cournapeau, D., Burovski, E., Peterson, P., Weckesser, W., Bright, J.,
van der Walt, S. J., Brett, M., Wilson, J., Millman, K. J., Mayorov, N.,
Nelson, A. R. J., Jones, E., Kern, R., Larson, E., \ldots{} SciPy 1. 0
Contributors. (2020). {SciPy 1.0: fundamental algorithms for scientific
computing in Python}. \emph{Nature Methods}, \emph{17}, 261--272.
\url{https://doi.org/10.1038/s41592-019-0686-2}

\end{CSLReferences}

\end{document}